\documentclass[aps,prb,twocolumn]{revtex4}
\usepackage{graphicx,epsf}
\usepackage{amsmath}

\newcommand{\ybco}{YBa$_2$Cu$_3$O$_{6+\delta}$}

\newcommand{\lsco}{La$_{2-x}$Sr$_x$CuO$_4$}
\newcommand{\bscco}{Bi$_2$Sr$_2$CaCu$_2$O$_{8+\delta}$}
\newcommand{\lbco}{La$_{2-x}$Ba$_x$CuO$_4$}
\newcommand{\lbcoo}{La$_{15/8}$Ba$_{1/8}$CuO$_4$}
\newcommand{\ccoc}{Ca$_{2-x}$Na$_x$CuO$_2$Cl$_2$}

\begin{document}

\title{
Superconducting $d$-wave stripes in cuprates: \\
Valence bond order coexisting with nodal quasiparticles
}

\author{Matthias Vojta}
\author{Oliver R\"osch}
\affiliation{Institut f\"ur Theoretische Physik, Universit\"at zu K\"oln,
Z\"ulpicher Str. 77, 50937 K\"oln, Germany}
\date{Jan 10, 2008}

\begin{abstract}
We point out that unidirectional bond-centered charge-density-wave states in cuprates
involve electronic order in both $s$- and $d$-wave channels,
with non-local Coulomb repulsion suppressing the $s$-wave component.
The resulting bond-charge-density wave, coexisting with
superconductivity, is compatible with recent photoemission and tunneling data
and as well as neutron-scattering measurements,
once long-range order is destroyed by slow fluctuations or glassy disorder.
In particular, the real-space structure of $d$-wave stripes is consistent with
the scanning-tunneling-microscopy measurements on both underdoped \bscco\ and \ccoc\
of Kohsaka {\em et al.} [Science {\bf 315}, 1380 (2007)].
\end{abstract}

\maketitle


\section{Introduction}

A remarkable aspect of the copper-oxide high-$T_c$ superconductors is that
various ordering phenomena apparently compete,
including commensurate and incommensurate magnetism,
superconducting pairing, and charge-density-wave formation.
(More exotic states have also been proposed, but not verified
experimentally beyond doubt.)
While commensurate magnetism and superconductivity are common
phases in essentially all families of cuprates, the role
of other instabilities for the global features of the phase
diagram is less clear.

A particularly interesting role is taken by charge-density waves.
Such states break the discrete lattice translation symmetry, with
examples being stripe, checkerboard, and valence-bond order.
In the compounds \lbco\ and \lsco\ (with Nd or Eu co-doping)
evidence for stripe-like spin and charge modulations
with static long-range order were detected,\cite{lsco,waki,pnas,christensen}
in particular near 1/8th doping.
(This is supported e.g. by strong phonon anomalies seen in
neutron scattering experiments.\cite{reznik})
While in other cuprate families similar long-range order has
not been found, signatures of short-range charge order, likely pinned
by impurities, have been observed in scanning tunneling microscopy (STM)
on underdoped \bscco\cite{ali,STM_others,kohsaka,mcelroy} and
\ccoc.\cite{kohsaka,hana}
The low-energy electronic structure in the presence of charge order turns
out to be remarkable:
In \lbcoo, angle-resolved photoemission (ARPES) indicated
a quasiparticle (QP) gap with $d$-wave like form, i.e.,
charge order coexists with gapless (nodal) QP in $(1,1)$ direction
(while antinodal QP near $(0,\pi)$ are gapped).\cite{valla}
STM data on both underdoped \bscco\ and \ccoc\ show QP
interference arising from coherent low-energy states near the nodes,
whereas electronic states at higher energy and wavevectors close to the antinode
are dominated by the real-space modulation of the
short-range charge order.\cite{mcelroy,kohsaka,hana07}
This dichotomy in momentum space is has also been
found in ARPES experiments in \lsco,\cite{zhou} \bscco,\cite{kanigel}
and \ccoc\cite{shen} where well-defined nodal and ill-defined
antinodal QP are frequently observed.

These results suggest that momentum-space differentiation and tendencies
toward charge ordering are common to underdoped cuprates.\cite{jan,ssrmp,stevek}
The concept of {\em fluctuating stripes}, i.e., almost charge-ordered
states, has been discussed early on.\cite{lsco,pnas,flstr,jan,stevek}
This concept, appropriate for compounds without static long-range order,
assumes the existence of a nearby stripe-ordered state,
with physical observables being influenced by the low-lying collective
modes associated with a charge-ordering instability.
Following this idea, we have recently calculated\cite{vvk} the spin excitation
spectrum of slowly fluctuating (or disordered) stripes.
We were able to show that fluctuating stripes give rise to an ``hour-glass'' magnetic
spectrum, very similar to that observed in neutron scattering experiments both
on \lbco\cite{jt04} and \ybco.\cite{hinkov,hayden}

\begin{figure}[!b]
\epsfxsize=2.5in
\epsffile{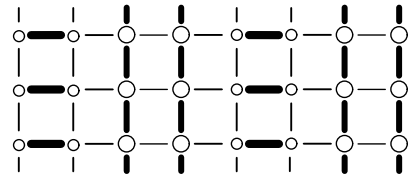}
\caption{
Schematic real-space structure of a stripe state with primarily
$d$-wave character and a $4\times 1$ unit cell, i.e. ${\bf Q}=(\pm\pi/2,0)$.
Cu lattice sites are shown as circles, with their size
representing the on-site hole densities. The line strengths indicate the
amplitude of bond variables like kinetic and magnetic energies.
The modulation 
in the site charge densities is small, whereas the one in the
bond densities is large and of $d$-wave type.\cite{dwstruct}
Note the similarity of the bond modulation with the STM data of
Ref.~\onlinecite{kohsaka}.
}
\label{fig:schem}
\end{figure}

The focus of this paper is on the electronic structure of stripe states.
We introduce the concept of ``$d$-wave stripes'':
Here the modulation of charge densities 
has primarily a $d$-wave form factor, i.e.,
lives more on the bonds than on the sites of the square lattice,
leaving nodal QP unaffected.
We illustrate that a picture of such bond-centered charge order,
coexisting with superconductivity
(this state may be dubbed ``valence-bond supersolid'')
is consistent with various features seen in both ARPES and STM
measurements.
In particular, the real-space pattern of $d$-wave stripes,
Fig.~\ref{fig:schem}, is strikingly similar to the STM results
of Ref.~\onlinecite{kohsaka}, obtained on underdoped \bscco\ and \ccoc.

Other types of $d$-wave particle--hole order have been discussed before.
(a) The $d$-density wave (or staggered-flux) phase\cite{ddw} was proposed as
a candidate for the pseudogap. However, it has no measurable charge modulation
and shall not be discussed here.
(b) Checkerboard (or plaquette) order\cite{dwave_checker1,dwave_checker2}
is related to stripes.
While both can occur as stable phases (with similar energetics) in variants of
the $t$-$J$ model,\cite{mv02,plaq2}
there are key experimental facts pointing towards stripe instead checkerboard
order being the primary instability:
(i) The STM data\cite{kohsaka} clearly show that the rotation symmetry is locally
broken from C$_4$ down to C$_2$.
(ii) The momentum-space pattern of spin excitations arising from checkerboard order
has been shown to be incompatible with the neutron response of materials like
\lbco\ or \ybco.\cite{carlson,vvk}


\section{Superconducting valence-bond states}

Numerous microscopic calculations, for Hubbard or $t$-$J$ models at low doping,
have indicated a tendency towards states with broken translational
symmetry.\cite{zaanen,schulz,machida,stevek,dmrg,doug,vs}
At low temperatures $T$, it is conceivable that this coexists with
superconductivity.\cite{vs,mv02,scstripes}
In fact, such a scenario can be expected on general grounds:
Upon destroying magnetic order in a square-lattice antiferromagnet (AF),
paramagnetic states with valence-bond (or spin-Peierls) order are known to appear.\cite{ssrmp}
The introduction of charge carriers by doping then leads
to superconductivity, coexisting with bond order for a finite
doping range.\cite{vs,jpsj}
A global phase diagram has been worked out using
a Sp($2N$) mean-field theory applied to the $t$-$J$ model supplemented by longer-range Coulomb
interaction:\cite{vs}
At small doping, superconducting bond-centered stripe states occur,
while homogeneous $d$-wave superconductivity is realized at larger doping.
(Depending on microscopic parameters, the stripes may get replaced by checkerboard
or spin-Peierls states.)
Related superconducting charge-ordered states also appear in other
theoretical treatments.\cite{scstripes,dwave_checker2}

Let us specify the various types of translational symmetry breaking on a square lattice,
assuming the magnetic SU(2) symmetry to be unbroken.
(a) Spin-Peierls states have a $2\times1$ unit cell where all sites are
equivalent, but the bonds are modulated. The C$_4$ rotation symmetry is broken,
the ordering wavevector is ${\bf Q}=(\pi,0)$,$(0,\pi)$.
(b) Stripe states have a $N\times1$ unit cell ($N=4$ is particularly stable),
both sites and bonds are modulated, ${\bf Q}=(\pm2\pi/N,0)$,$(0,\pm2\pi/N)$, and C$_4$ is broken.
(c) Checkerboard states have a $N\times N$ unit cell,
both sites and bonds are modulated, ${\bf Q}=(\pm2\pi/N,\pm2\pi/N)$, but C$_4$ is intact.
While stripe and checkerboard states can in principle be either site- or bond-centered,
experimental evidence points towards bond-centered structures.\cite{jt04,abbamonte,kohsaka}

In a quasiparticle picture, the symmetry-breaking orders can be translated into
expectation values of fermionic bilinears:
$\phi_1({\bf k}) = \langle c_{{\bf k}\uparrow} c_{-{\bf k}\downarrow} \rangle$
captures homogeneous superconductivity,
with $\phi_1({\bf k})\propto \cos k_x-\cos k_y$ for $d$-wave pairing,
while
$\phi_2({\bf k}) = \langle c_{{\bf k+Q},\sigma}^\dagger  c_{{\bf k}\sigma} \rangle$
and
$\phi_3({\bf k}) = \langle c_{{\bf k+Q},\uparrow} c_{-{\bf k}\downarrow} \rangle$
originate from charge order.
The directional dependence of $\phi_i({\bf k})$ can be decomposed according
to the representations of the point group.
In the presence of unidirectional order, i.e., C$_4$ broken down to C$_2$,
the $s$-wave and $d$-wave order parameters inevitably mix.
Our superconducting stripe states below will have $\phi_{1,2,3}$ all non-zero.
(Note that e.g. $\phi_3$ alone generates a Fulde-Ferrell-Larkin-Ovchinnikov state.)


\section{$d$-wave stripes}

Stripe states are best discussed in real space
(phrased in the following for a square lattice of Cu atoms, keeping
in mind that bonds of this lattice correspond to oxygen orbitals).
Naively, the primary phenomenon of stripe order is a modulation of the
on-site charge densities, $\langle c_{i\sigma}^\dagger c_{i\sigma}\rangle$,
which translates into a momentum-independent (i.e. $s$-wave) order parameter
$\phi_2({\bf k})$.
However, general arguments indicate\cite{vs,ssrmp} that the physics behind
local ordering acts on bonds instead of sites:
Stripe formation is driven by the competition between kinetic and magnetic
energies, both living on lattice bonds.
In such a bond-dominated stripe state,
modulations in quantities like $\langle c_{i\sigma}^\dagger c_{i+\Delta,\sigma}\rangle$
can locally have different signs on horizontal and vertical bonds,
implying a $d$-wave component of $\phi_2({\bf k})$.

Given the mixing of $s$- and $d$-wave components, stripe states may have
ordering primarily in either the $s$- or the $d$-wave channel.
In the following, we shall argue in favor of stripes dominated by the $d$-wave component.
Fourier-transforming $\phi_2({\bf k})$ into real space\cite{dwstruct}
leads to a stripe state as in Fig.~\ref{fig:schem} --
this is one of the main results of this paper.
Importantly, the ordering pattern in Fig.~\ref{fig:schem} appears perfectly consistent
with the STM data of Kohsaka {\em et al.} (Fig. 4 of Ref.~\onlinecite{kohsaka}),
where locally well-formed period-4 structures are seen,
consisting of distinct ladder-like objects.

Arguments in favor of stripes with a dominant $d$-wave component (Fig.~\ref{fig:schem})
are:\cite{honer}
(i) Energetics:
In the hole-poor regions, strong horizontal bonds form due to the tendency
towards dimerization (i.e. optimizing the magnetic energy),
whereas in the hole-rich regions strong vertical bonds form optimizing
the hole kinetic energy.
(ii) Coulomb interaction disfavors on-site charge modulations and
thus suppresses the $s$-wave component.

If stripe order coexists with superconductivity, then the bond pairing
amplitudes, $\langle c_{i\uparrow} c_{i+\Delta,\downarrow}\rangle$,
will be non-zero as well.
Starting from a homogeneous $d$-wave superconductor,
stripe order will induce a modulated $s$-wave pairing component, described
by $\phi_3({\bf k})$.


\section{Mean-field theory}

Superconducting stripe states can be obtained in
Sp($2N$) mean-field theory.\cite{vs}
Consider an extended $t$-$J$ Hamiltonian for
fermions without double occupancies, $c_{i \alpha}$, with spin $\alpha=1 \ldots 2N$
($N=1$ is the physical value):
\begin{eqnarray}
{\cal H}_{tJV} = \sum_{i > j} && \left[ -\frac{t_{ij}}{N} \sum_\alpha c_{i \alpha}^{\dagger} c_{j
\alpha} + {\rm H.c.}  + \frac{V_{ij}}{N} n_i n_j \right. \nonumber \\
&&~~~~~~~~+\left. \frac{J_{ij}}{N} \left( {\bf S}_i \cdot {\bf S}_j -
\frac{n_i n_j}{4N} \right) \right]
\label{e1}
\end{eqnarray}
with $n_i = \sum_\alpha c_{i \alpha}^{\dagger} c_{i \alpha}$.
The spin operators ${\bf S}_i$ are fermion
bilinears times the traceless generators of ${\rm Sp} (2N)$.
The fermion hopping, $t_{ij}$, will be non-zero for nearest and next-nearest neighbors,
with values $t$ and $t'$, while the exchange, $J_{ij}$, is restricted to nearest-neighbor
terms, $J$. The average doping $\delta$ is fixed by
$\sum_i \langle n_i \rangle = N N_s (1-\delta)$,
where $N_s$ is the number of lattice sites.

\begin{figure}[!t]
\epsfxsize=2.8in
\epsffile{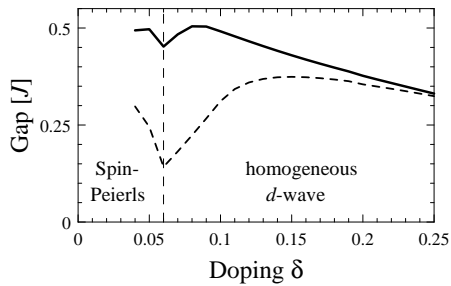}
\caption{
Doping dependence of quasiparticle gap energies from Sp($2N$) mean-field theory.
The solid line shows the antinodal gap, i.e., the minimum QP energy along $(k_x,\pi)$,
for $t/J=4$, $t'/t=-0.25$ -- here only spin-Peierls order, but no stripes occur
for doping $\delta>3\%$.
(For stripe states, the characteristic gap scale is more difficult to extract
due to band folding.)
In addition, the dashed line shows the minimum QP energy along the boundary of
the AF Brillouin zone (BZ).\cite{jcdpriv}
}
\label{fig:gap}
\end{figure}

The spins and holes can be represented by auxiliary fermions $f_{i\alpha}$
and spinless bosons $b_i$, respectively, such that the physical
electrons $c_{i\alpha} = b_i^\dagger f_{i\alpha}$.
Via a Hubbard-Stratonovich decoupling of the AF
interaction we introduce link fields $Q_{ij}$, defined on the
bonds of the square lattice.
After taking the limit $N\to\infty$, the slave bosons $b_i$ condense,
$\langle b_i\rangle = \sqrt{N}b_i$, and
the $Q_{ij}$ take static saddle-point values.
We are left with a bilinear Hamiltonian which can be diagonalized by
a Bogoliubov transformation.
At the saddle point, the slave-boson amplitudes fulfill
$\sum_i b_i^2 = N_s \delta$, and the link fields are given by
$N Q_{ij}=\langle {\cal J}^{\alpha\beta} f_{i\alpha}^\dagger f_{j\beta}^\dagger \rangle$,
where ${\cal J}^{\alpha\beta}$ is the antisymmetric ${\rm Sp}(2N)$ tensor.
The distribution of the $b_i$ is found be minimizing the saddle-point
free energy, here the Coulomb repulsion $V_{ij}$ enters
(on a classical level only).


At non-zero doping and $T=0$, all mean-field phases are superconducting.
At larger $t/J$, the on-site charge distribution is homogeneous except for tiny doping,
but bond-order of spin-Peierls type occurs at low doping.
The doping dependence of the gap energy scale is shown in Fig.~\ref{fig:gap};
it roughly follows the experimentally established pseudogap scale $T^\ast$.

At smaller $t/J$, stripe order occurs over a sizable doping range.
As in most mean-field theories,
the tendency toward ordering is overestimated:
In a large parameter regime, the mean-field theory predicts bond-centered
stripes with maximal charge inhomogeneity and strong bond order.\cite{vs}
To obtain the electronic structure for a more realistic stripe state from
the mean-field theory, we have employed the following modifications:
(i) The distribution of on-site hole densities within the unit cell
is enforced by hand through the $b_i$ values.
(ii) The parameter $t/J$ is chosen such that the system is in the
bond-ordered regime, but close to the transition to homogeneous
superconductivity.
These modifications anticipate that both quantum effects
and Coulomb repulsion (beyond the classical approximation) reduce the
amplitude of the charge modulations.\cite{beyond}
[Density-matrix renormalization group calculations
for the 2d $t$-$J$ model (with realistic $t'$) showed signatures for bond-centered
stripes with reduced on-site modulation amplitude.\cite{dmrg,doug}]

\begin{figure}[!t]
\epsfxsize=3.0in
\epsffile{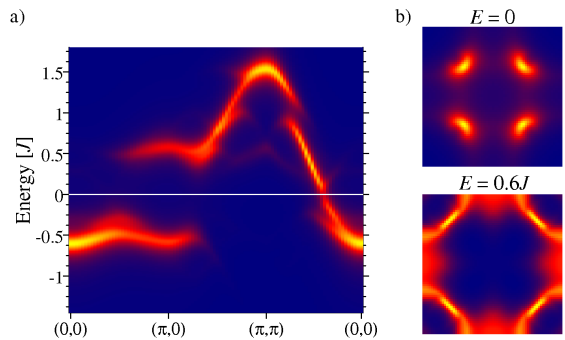}
\caption{
Electronic structure of a $4\times 1$ stripe state from
mean-field theory, here with $t/J=1.95$, $t'/t=-0.25$,
overall doping $\delta=1/8$, and a
hole distribution of $1.3 : 0.7$ enforced by hand,
see text.
a) Dispersion along high-symmetry lines in $k$ space.
b) Constant-energy cuts through $k$ space, showing nodal and antinodal structures.
The spectra of horizontal and vertical stripes have been added,
and an artificial broadening of $J/20$ was used.
}
\label{fig:dsp}
\end{figure}

A sample result for the electronic spectrum of a period-4 stripe
state is shown in Fig.~\ref{fig:dsp}.
In this calculation, the hole distribution was fairly inhomogeneous,
$b_1^2=b_2^2 = 1.3 \delta$, $b_3^2=b_4^2 = 0.7 \delta$,
and the link fields $|Q_{ij}|$ varied by a factor of four within the unit cell.
Thus, the stripe state contained both sizable $s$- and $d$-wave components.
Nevertheless, the spectrum is essentially gapless along the $k$-space
diagonal, while a large gap appears near the antinodes, Fig.~\ref{fig:dsp}.
This toy calculation illustrates a key point, namely bond-centered stripe order
{\em is compatible} with the presence of nodal quasiparticles,
provided that the $s$-wave component of the order is not too
large\cite{granath,wavev} -- this compatibility was subject of discussions
in the past.\cite{palee}

\section{Glassy disorder}

Let us comment on the influence of quenched disorder arising from dopant impurities.
In situations where a symmetry-breaking order is only ``almost'' static,
adding disorder can pin the fluctuations and locally stabilize ordered islands.
The result is a state with static short-range order, not unlike in glassy systems.
For stripe order, this means that domains with segments of horizontal and vertical stripes
will coexist, with a checkerboard structure of domain walls.\cite{vvk,maestro}
The electronic QP excitations now move in a static disorder potential,
which couples strongly only to the QP near the antinodes,
while nodal QP are little affected.
As a result, the scattering rate along the normal-state Fermi surface
will be strongly energy-dependent, and gapless coherent nodal QP will
coexist with incoherent antinodal QP.

Importantly, the disorder potential acts as a random {\em field} on
charge stripes, smearing out any finite-temperature phase transition
in the charge sector.\cite{stevek,maestro}
(This is distinct from the spin sector, where static order at low $T$ is
still accompanied by a sharp phase transition, as the disorder
is of random-mass type.)
At elevated temperatures, signatures of local valence-bond (i.e. singlet)
formation will be visible below a temperature $T^\ast \sim J$,
and will likely evolve continuously from the high-temperature ``pseudogap''
scale to the low-temperature ``superconducting'' gap.


\section{Theory vs. Experiment}

The described picture of valence-bond order, coexisting with superconductivity at low $T$,
nicely ties in with various features of recent experimental data.
(i) Models of coupled spin ladders, arising from bond-centered stripes,\cite{MVTU,GSU}
provide an excellent description of the spin dynamics in stripe-ordered \lbcoo.\cite{jt04}
(ii) The short-range charge order seen in STM on underdoped \bscco\ and \ccoc\cite{kohsaka}
is bond-centered and locally breaks the lattice rotation symmetry down to C$_2$
The glassy real-space structure is compatible with impurity-pinned short-range
order, where fluctuations in the charge order parameter are primarily of
phase instead of amplitude type.\cite{vvk}
(iii) The apparent $d$-wave gap of the non-superconducting stripe compound
\lbcoo\ is naturally explained by $d$-wave stripes.
(iv) To explain the absence of 3d superconductivity in \lbcoo,
``antiphase superconductivity'' has been proposed.\cite{antiphase}
This is similar to $\phi_3({\bf k})$ above, and broadly consistent with
the picture advocated here.


\section{Conclusions}

Building on earlier work on charge order in cuprates,
we have pointed out that bond-ordered stripe states, possibly of
glassy character and coexisting with superconductivity,
provide a phenomenological framework which is consistent with
many features of recent cuprate experiments.
We believe that this strengthens the case for bond order (and associated
local singlet formation) to be a common tendency in cuprates, likely relevant also
for pseudogap phenomena.
Let us note, however, that we think that stripe-like translational symmetry
breaking is a phenomenon {\em competing} with superconductivity,
i.e., pairing is suppressed by stripes with spatial long-range order.


\acknowledgments

We thank J. C. S. Davis, C. Honerkamp, B. Keimer, H. Monien, H. Takagi,
and A. Yazdani for discussions,
and especially S. Sachdev for collaborations on related work.
This research was supported by the DFG through
SFB 608 (K\"oln) and the Research Unit FG 538.


\end{document}